
%
%
%
%
%
%
%

\def\unlock{
 \catcode`@=11 }
\unlock

 \font\fourteenrm=cmr10                 scaled\magstep2
 \font\twelverm=cmr10                   scaled\magstep1
 \font\ninerm=cmr9            \font\sixrm=cmr6

 \font\fourteenbf=cmbx10                scaled\magstep2
 \font\twelvebf=cmbx10                  scaled\magstep1
 \font\ninebf=cmbx9            \font\sixbf=cmbx5
 \font\twentyfouri=cmmi10 scaled\magstep4  \skewchar\twentyfouri='177
 \font\seventeeni=cmmi10  scaled\magstep3  \skewchar\seventeeni='177
 \font\fourteeni=cmmi10   scaled\magstep2  \skewchar\fourteeni='177
 \font\twelvei=cmmi10     scaled\magstep1  \skewchar\twelvei='177
 \font\ninei=cmmi9                         \skewchar\ninei='177
 \font\sixi=cmmi6                          \skewchar\sixi='177
 \font\twentyfoursy=cmsy10 scaled\magstep4 \skewchar\twentyfoursy='60
 \font\seventeensy=cmsy10  scaled\magstep3 \skewchar\seventeensy='60
 \font\fourteensy=cmsy10   scaled\magstep2 \skewchar\fourteensy='60
 \font\twelvesy=cmsy10     scaled\magstep1 \skewchar\twelvesy='60
 \font\ninesy=cmsy9                        \skewchar\ninesy='60
 \font\sixsy=cmsy6                         \skewchar\sixsy='60

 \font\fourteenex=cmex10    scaled\magstep2
 \font\twelveex=cmex10      scaled\magstep1
 \font\elevenex=cmex10      scaled\magstephalf
 \font\tenex=cmex10
 \font\nineex=cmex10 at 9pt

 \font\fourteensl=cmsl10    scaled\magstep2
 \font\twelvesl=cmsl10      scaled\magstep1
 \font\ninesl=cmsl10 at 9pt

 \font\fourteenit=cmti10   scaled\magstep2
 \font\twelveit=cmti10     scaled\magstep1
 \font\tenit=cmti10
 \font\nineit=cmti10 at 9pt
 
 \font\twelvett=cmtt10     scaled\magstep1
 \font\tentt=cmtt10
 \font\ninett=cmtt10 at 9pt
 
 \font\twelvecp=cmcsc10    scaled\magstep1
 \font\tencp=cmcsc10
 \font\ninecp=cmcsc10 at 9pt

 \newfam\cpfam
 \newcount\f@ntkey            \f@ntkey=0
 \def\samef@nt{\relax \ifcase\f@ntkey \rm \or\oldstyle \or\or
          \or\it \or\sl \or\bf \or\tt \or\caps \fi }

 \def\fourteenpoint{\relax
     \textfont0=\fourteenrm
     \scriptfont0=\tenrm
     \scriptscriptfont0=\sevenrm
      \def\rm{\fam0 \fourteenrm \f@ntkey=0 }\relax
     \textfont1=\fourteeni           \scriptfont1=\teni
     \scriptscriptfont1=\seveni
      \def\oldstyle{\fam1 \fourteeni\f@ntkey=1 }\relax
     \textfont2=\fourteensy          \scriptfont2=\tensy
     \scriptscriptfont2=\sevensy
     \textfont3=\fourteenex          \scriptfont3=\twelveex
     \scriptscriptfont3=\tenex
     \def\it{\fam\itfam \fourteenit\f@ntkey=4 }\textfont\itfam=\fourteenit
     \def\sl{\fam\slfam \fourteensl\f@ntkey=5 }\textfont\slfam=\fourteensl
     \scriptfont\slfam=\tensl
     \def\bf{\fam\bffam \fourteenbf\f@ntkey=6 }\textfont\bffam=\fourteenbf
     \scriptfont\bffam=\tenbf     \scriptscriptfont\bffam=\sevenbf
     \def\tt{\fam\ttfam \twelvett \f@ntkey=7 }\textfont\ttfam=\twelvett
     \def\caps{\fam\cpfam \twelvecp \f@ntkey=8 }\textfont\cpfam=\twelvecp
   \samef@nt
 }

 \def\twelvepoint{\relax
     \textfont0=\twelverm
      \scriptfont0=\ninerm
     \scriptscriptfont0=\sixrm
     \def\rm{\fam0 \twelverm \f@ntkey=0 }\relax
     \textfont1=\twelvei           \scriptfont1=\ninei
     \scriptscriptfont1=\sixi
      \def\oldstyle{\fam1 \twelvei\f@ntkey=1 }\relax
     \textfont2=\twelvesy          \scriptfont2=\ninesy
     \scriptscriptfont2=\sixsy
     \textfont3=\twelveex          \scriptfont3=\elevenex
     \scriptscriptfont3=\tenex
     \def\it{\fam\itfam \twelveit \f@ntkey=4 }\textfont\itfam=\twelveit
     \def\sl{\fam\slfam \twelvesl \f@ntkey=5 }\textfont\slfam=\twelvesl
     \scriptfont\slfam=\ninesl
     \def\bf{\fam\bffam \twelvebf \f@ntkey=6 }\textfont\bffam=\twelvebf
     \scriptfont\bffam=\ninebf     \scriptscriptfont\bffam=\sixbf
     \def\tt{\fam\ttfam \twelvett \f@ntkey=7 }\textfont\ttfam=\twelvett
     \def\caps{\fam\cpfam \twelvecp \f@ntkey=8 }\textfont\cpfam=\twelvecp
     \samef@nt
 }

 \def\tenpoint{\relax
     \textfont0=\tenrm          \scriptfont0=\sevenrm
     \scriptscriptfont0=\fiverm
     \def\rm{\fam0 \tenrm \f@ntkey=0 }\relax
     \textfont1=\teni           \scriptfont1=\seveni
     \scriptscriptfont1=\fivei
     \def\oldstyle{\fam1 \teni \f@ntkey=1 }\relax
     \textfont2=\tensy          \scriptfont2=\sevensy
     \scriptscriptfont2=\fivesy
     \textfont3=\tenex          \scriptfont3=\tenex
     \scriptscriptfont3=\tenex
     \def\it{\fam\itfam \tenit \f@ntkey=4 }\textfont\itfam=\tenit
     \def\sl{\fam\slfam \tensl \f@ntkey=5 }\textfont\slfam=\tensl
     \def\bf{\fam\bffam \tenbf \f@ntkey=6 }\textfont\bffam=\tenbf
     \scriptfont\bffam=\sevenbf     \scriptscriptfont\bffam=\fivebf
     \def\tt{\fam\ttfam \tentt \f@ntkey=7 }\textfont\ttfam=\tentt
     \def\caps{\fam\cpfam \tencp \f@ntkey=8 }\textfont\cpfam=\tencp
     \setbox\strutbox=\hbox{\vrule height 8.5pt depth 3.5pt width\z@}
     \samef@nt}

 \def\ninepoint{\relax
     \textfont0=\ninerm          \scriptfont0=\sevenrm
     \scriptscriptfont0=\fiverm
     \def\rm{\fam0 \ninerm \f@ntkey=0 }\relax
     \textfont1=\ninei           \scriptfont1=\seveni
     \scriptscriptfont1=\fivei
     \def\oldstyle{\fam1 \ninei \f@ntkey=1 }\relax
     \textfont2=\ninesy          \scriptfont2=\sevensy
     \scriptscriptfont2=\fivesy
     \textfont3=\nineex          \scriptfont3=\nineex
     \scriptscriptfont3=\nineex
     \def\it{\fam\itfam \nineit \f@ntkey=4 }\textfont\itfam=\nineit
     \def\sl{\fam\slfam \ninesl \f@ntkey=5 }\textfont\slfam=\ninesl
     \def\bf{\fam\bffam \ninebf \f@ntkey=6 }\textfont\bffam=\ninebf
     \scriptfont\bffam=\sevenbf     \scriptscriptfont\bffam=\fivebf
     \def\tt{\fam\ttfam \ninett \f@ntkey=7 }\textfont\ttfam=\ninett
     \def\caps{\fam\cpfam \ninecp \f@ntkey=8 }\textfont\cpfam=\ninecp
     \setbox\strutbox=\hbox{\vrule height 8.5pt depth 3.5pt width\z@}
     \samef@nt}

\newcount\sectionnumber    \sectionnumber=0
\def\section#1{\goodbreak
\vskip 0.8cm
 \global\advance\sectionnumber by 1
\line{\bf\the\sectionnumber . #1\hfil}
\vskip 0.4cm}

\def\references{\goodbreak
\vskip 0.8cm
\line{\bf References.\hfil}
\vskip 0.4cm}

\def\abstract#1{\centerline{\tenrm ABSTRACT}
\vglue 0.3cm \rightskip=3pc \leftskip=3pc
\tenrm\baselineskip=12pt \parindent=1pc
 #1 \vglue 0.8cm }

%
\def\labstract#1{\centerline{\tenpoint\rm ABSTRACT}
\centerline{\tenpoint\rm \noindent #1}
\vglue 0.3cm }

\parindent=3pc
\baselineskip=10pt
\hsize=6.0truein
\vsize=8.5truein

\footline{\rm\hss\folio\hss}
\line{\hfil }
\vglue 0.5cm

\twelvepoint

\setbox0=\vbox{\hsize=40mm \ninepoint
\noindent Submitted for publication in Acta Physica Polonica.}

\line{\box0 \hfil UM-TH-94-12}
\vskip 0.8cm

\centerline{\tenpoint\bf PERTURBATION THEORY
 AND RELATIVE SPACE\footnote{\tenpoint\rm
$^\dagger$}{\ninepoint Presented at the conference
 ``Hard Problems in Mathematical Physics'', at the occasion of
 the Sixtieth Birthday of Paul Federbush. Ann Arbor, May 2-4 1994.}}

\vglue 1.0cm
\centerline{\tenpoint M. VELTMAN}
\baselineskip=13pt plus  0.2pt
\centerline{\tenpoint\it Department of Physics, University of Michigan}
\baselineskip=12pt plus 1pt
\centerline{\tenpoint\it Ann Arbor, Michigan 48109, USA}
\vglue 0.8cm

\labstract{The validity of non-perturbative methods is questioned.
The concept of relative space is introduced.}

\baselineskip=14pt plus 0.2pt
\twelvepoint

\section{Introduction.}
Both success and failure have been spectacular in present day field
theory and particle physics. High energy experiments, so far, confirm
the Standard Model to an astonishing degree. That same model has many
arbitrary features; all attempts at understanding these features have
failed abysmally. The origin of the particular symmetries of the
Standard Model, SU(3)$\times$SU(2)$\times$U(1) is unknown. We have no
idea why there are three generations of particles, nor do we have a
clue as to the particular multiplets chosen by nature. Coupling
constants and masses of the model are unexplained. And so on.

The only theoretical successes in explaining the Standard Model are
certain consequences of the SU(5) grand unification scheme. That theory
provided us with numerical values for the bottom quark mass and the weak
mixing angle, with reasonable agreement with experiment. Furthermore,
massless neutrino's are natural in that scheme. At some point the model
fails, and it must also not be overlooked that it is at most a very
partial solution to our problems: it leaves most of the basic questions
unanswered.

The idea of renormalizable field theory, so succesful in its application
to the Standard Model, has resulted in further theoretical advances,
notably supergravity and string theory. Unfortunately, not one single
question of the type cited above has been answered, and the theories
show unmistakably signs of that same old malady that we will lump under
the name ``epicycles''.

The concept of naturalness is usually cited as the underlying motivation
for supersymmetry. We will challenge that concept, and in any case need
to point out that there is nothing natural about the development of the
theory itself. Its main success is its agility in dodging the facts. The
dubious explanation of the convergence of the three scaling coupling
constants into a single point can not be taken seriously. It is just
another fit, using some of the many free parameters.

The list of failures is as long as the list of attempts, and it is
pointless to discuss the often extremely ingenious constructs. However,
it must be noted that there are some real difficulties in the Standard
Model, quite apart from questions concerning the origin of the
particular features of that model. The most pronounced of these concern
the cosmological constant and strong $CP$ violation. Despite great
theoretical and experimental efforts (axions) no reasonable solution to
these problems has been offered. More in general, failure seems
automatic whenever non-perturbative aspects of the Standard Model are
considered.

In the background, as always, lurks non-renormalizable gravitation with
its black and other holes. The Higgs system and the associated problem
of the cosmological constant evokes the impression that there is a deep
and fundamental connection that so far remains completely hidden to us.
It certainly appears that the problem is too difficult for us, and very
likely, only experiment can help us to gain insight.

A re-examination of some very basic concepts appears timely and
expedient. We must clarify the presently achieved description of nature
for the simple reason that there is more than one viewpoint.
There are different formalisms, presumably describing physical
reality, but it is not clear that the descriptions are actually
equivalent.

\section{Complementarity.}
This century has seen the introduction of two great theoretical
creations, relativity and quantum mechanics. Almost from the start,
conflict has surrounded the meeting of the two, and in fact persists to
this day. It must be understood that the problem of the cosmological
constant is precisely a consequence of the basic features of both
theories. Einstein's theory of gravitation, by necessity, introduces
the metric of the underlying space as a free parameter, the
cosmological constant. Within the context of classical general
relativity that constant may be chosen to be zero, even if this appears
to us today as an arbitrary choice. Quantum mechanics however radically
changes the situation, simply because it affects that constant. If it
is initially chosen to be zero radiative corrections will change that.
The present day observation of a very small if not zero cosmological
constant is in flagrant contradiction with the scale of the corrections
suggested by quantum theory.

In first instance, in the twenties, the conflict manifested itself
through the famous discussions between Bohr and Einstein. While on the
face of it Bohr appeared to have the upper hand, it is nonetheless
clear that there is more to the issue. Bell's inequalities and the
associated literature testify to that. Einstein, to the very end, has
refused to accept the quantum concepts, and it may well be that he has
perceived the fundamental conflict more clearly than anyone else.

In the Copenhagen philosophy there is the concept of complementary,
which we will interpret to mean the following. There are two alternative
descriptions of particles, namely the particle description and the wave
description. The particle description is by means of momentum and
energy, the wave description concerns location in time and space.
Basically, one is the Fourier transform of the other, and we could
simply state that either description is complete and supposedly fully
equivalent to the other. A particle may be specified by a superposition
of momentum states or by a wave function in space-time (coordinate
space). Either space, momentum space or coordinate space, may be used to
describe the situation. We may use this or that representation in
setting up Hilbert space.

But is it really true that momentum space and coordinate space are
equivalent? Here is the basic conflict: in Einstein's concept of
gravitation they are not possibly equivalent. In general relativity
space-time plays a very particular role, it is intimately connected with
gravitation. Gravitation is interpreted as nothing else but the
structure of space-time. In turn, that structure is determined by the
distributed matter.

Why then, if the description in momentum space is equivalent, do we not
introduce a metric in momentum space? Is that space flat by definition?
Putting the issue this way the basic conflict between gravitation and
quantum mechanics becomes obvious. Gravitation is particular to
space-time. By its very nature general relativity assigns properties to
space-time. But it is completely unclear whether a definition of physics
in momentum space would adhere to these assumed properties of space
time. It may conflict.

In order to investigate this question we must clarify our description of
physical reality. We can not, on the one hand, do gauge theories and
renormalizable field theory in momentum space, and on the other hand
solve classical equations of motion and play with black holes in
coordinate space. In other words, we must realize that these descriptions
may not be equivalent, and that we may have to make a choice depending
on the agreement with observed physics.

\section{Theoretical Framework.}
For the moment we will leave gravitation and concentrate on quantum
field theory. Here, today, we have three seemingly equivalent
descriptions. They are:

 \item{-} The canonical formalism, involving Lagrangian, coordinates and
their conjugate momenta, and an $S$-matrix defined in terms of
time-ordered products of operators in Hilbert space.

 \item{-} The path integral formalism where the $S$-matrix is defined as
a sum over paths in coordinate space.

 \item{-} The purely pragmatic description of the $S$-matrix in terms of
Feynman rules with in addition the prescriptions of dimensional
regularization. We will call this the dimensional formulation.

\noindent The third prescription is totally perturbative, but otherwise
complete in itself. Also, it may assure us with respect to anomalies. Both
the canonical and path integral formalism are at least in principle not
restricted to the perturbative domain, but on the other hand they need
in addition a regulator method. Usually somewhere along the line
dimensional regularization is used, but it must be stated that that
method can not be formulated within either canonical or path integral
formalism. The conventional wisdom is that one can use any regulator
method desired and that with the appropriate counter terms the results
are independent of the particular regulator method used. Ward identities
are the crucial instruments. Then one may use any method that can be
formulated in either scheme, be it Pauli-Villars regularization, or a
finite lattice spacing, or whatever.

The crucial question is whether the conventional wisdom is correct. The
answer is that it appears correct, but only within the context of
renormalizable perturbation theory, that is within the domain of the
dimensional formulation.

The question of regularization scheme is an old one, and we may perhaps
briefly reflect on that. Discovering the basic difficulty of infinities
in field theory, Lorentz speculated that perhaps the electron has an
extended structure in space-time. The idea of renormalization is that
many physical properties do not depend on the details of that structure,
and that the only consequence is a redefinition of the mass of the
electron. However, it must be realized that this idea has become totally
untenable in modern gauge field theory. The problem is that any attempt
at regularization through a finite extension in space-time hopelessly
conflicts with gauge invariance. To make it explicit, suppose the
electron coupling to vector bosons involves a form factor (the Fourier
transform of its spatial structure). That, through Ward identities for
the case of vector boson electron scattering, has its implications for
the three vector boson vertex, but it is not obvious that these
implications can be put in terms of a form factor for the three vector
boson vertex (it can not). And Ward identities for vector boson
scattering, that involve also the four point vector boson vertex, are
conflicting with the assumption of form factors for the vertices
involved. In other words, the intuitive idea that a finite space-time
structure of particles would ultimately solve the problems of infinities
in field theory is removed farther than ever from realization.

Another possible point of view is that all particles are basically
massless and acquire mass through an essentially low energy mechanism.
The divergence structure of the theory is then the divergence structure
of a massless theory. No one has achieved to regulate that in a
physically appealing way. It is interesting to note in this context that
in the Standard Model indeed all particles, fermions as well as vector
bosons, acquire their mass through the Higgs mechanism. The way the Higgs
boson itself acquires mass is less clear.

It should be added that the renormalization prescription has become much
more decoupled from the physical situation. While in Lorentz's view all
difficulties could be concentrated in a simple physical picture, that of
an extended electron, present day renormalization is much less directly
related to any physical image. The renormalization prescription is now
simply a matter of fixing parameters, accidentally involving infinities,
and no one associates that with any particular physical visualization.
Again, a view based on a particular perspective in coordinate space
fails, in fact is contradictory to the theory.

Because of these considerations one may feel inclined to break with all
these vague hopes and insights. As a possible alternative to Hamiltonian
formalism or path integrals we therefore bluntly take as starting point
perturbative field theory, defined in terms of Feynman diagrams, using
dimensional regularization, i.e.~the dimensional formulation. Thus loop
momenta are in $n$-dimensional space, and at the end the limit $n=4$
must be taken. Fitting the free parameters to the experiment provides
excellent agreement with experiment, from Coulomb law via Lamb shift to
LEP observations.

At this point it must be mentioned that the dimensional formulation
is not free of problems either. There are situations where perturbation
theory is non-convergent, and it is not clear how to deal with that.
A point in case is the non-convergence of quantum chromodynamics in the
infrared. This is usually referred to as the confinement problem, a very
specific coordinate space type qualification. Indeed, in momentum space
one has a problem. But let us not forget that no solution of the
confinement problem has been offered so far. Perhaps a non-coordinate
point of view is more productive.

\section{Space-time.}
In the dimensional formulation no assumptions concerning space and time
occur, they do not occur altogether. This is the central point. Space
and time do not occur in any way in this definition of physical reality.
We repeat and emphasize: space and time do not occur in the dimensional
definition of physical reality.

It follows that behaviour in space-time is solely defined by Fourier
transformation. It must be understood that the Fourier transform of
$n$-dimensional momentum space (with continuous $n$) is not simply
$n$-dimensional coordinate space. It is unclear how to define the
Fourier transform. Obviously then the dimensional formulation is in
contradiction with the assumption of four dimensional coordinate space.
Whether in the limit $n=4$ the conflict resolves is another matter.
However, this formulation of physical reality is not equivalent to
canonical or path integral formulations. As argued above, these latter
formalisms need dimensional regularization to show their consistency,
obviously restricted to the perturbative regime. In other words, the
other formalisms are well defined insofar they are perturbative. We do
not know to what extent non-perturbative results are true.

In the dimensional formulation, therefore, space and time have no a
priori existence. They exist exclusively as Fourier transforms of a
momentum space description. Space and time are defined solely relative
to momentum space. To what extent absolute properties can be ascribed to
momentum space is not our concern here, nor is there any need. But we
must not assign absolute properties to space-time as they may simply
conflict with the starting point. For example, a cosmological constant
assigns a metric to space-time as a boundary condition, in the absence
of matter. However, how can we assign a metric to a space defined
through Fourier transforms? What is a boundary condition in a Fourier
transformed space? To make it very explicit, consider the idea that one
would impose a boundary condition in momentum space, or define a metric
in that space. It just makes no sense.

Within the dimensional scheme we must abandon the idea of the absolute
existence of a space-time continuum. There are arguments leading up to
the idea of an absolute time continuum, but this is not the moment to
discuss that. The statement that if a particle moves from one point to
another, it must necessarily pass through some series of points
in between is meaningless. Of course, some kind of continuity property
can certainly be derived, but it is not true a priory. It is not an
intrinsic property of space-time, but at best a derived property.

The idea of a vacuum expectation value never really occurs in the
dimensional formulation. We simply start with Feynman rules that
correspond to a Lagrangian with the Higgs field shifted by a constant.
The issue surfaces only when discussing gravitation and the cosmological
constant.

The battered concept of causality loses much of its meaning in the
dimensional viewpoint described here. Unlike unitarity it becomes a
derived property, no fundamental assumption. The mathematical properties
that guarantee unitarity actually largely imply locality as well. But
then, who cares about causality in some mathematical space? Indeed,
quantum mechanics has always been very ambiguous on this point, and the
situation becomes intolerable when gravitation and its black holes enter
the discussion. Recent arguments in the literature point to drastic
difficulties.

It is curious to note that Einstein was well aware of the concept of
absolute space-time in relation to his theory. In certain editions of his
book$^1$ ``The Meaning of Relativity'' he makes some general remarks on the
issue. The discussion is very interesting, but any partial quote would
do unjustice to the whole and we leave it at this. It appears
quite possible that Einstein suspected that his philosophy was at odds
with the ideas of quantum mechanics. That could explain his reluctance
towards accepting quantum theory.

It is perhaps necessary at this point to state explicitly that most
consequences of Einstein's theory remain true, also in the dimensional
formulation. The philosophy changes, and furthermore cosmology needs
re-examination. Black holes are probably nothing else but commercially
viable figments of the imagination.
Unfortunately, Einstein's beautiful dream, to formulate forces as
properties of space-time, has already been next to untenable for quite
some time, given the multitude of forces that we have to deal with.
Few would reject the dimensional formulation because it is at odds
with a geometric interpretation.

The point of view arising from the dimensional formulation is utterly
alien to our usual intuition concerning space-time. Yet there is no
logical reason that can be put against it; moreover the usual troubles
of quantum mechanics, such as wave function collapse, Bell's
inequalities etc.~rather obviously point in the direction described.
And let us not forget the old ugly aspect of curved space: general
coordinate transformations have no half integer spin representations.
That beauty defect disappears. Momentum space is perfectly flat. Gauge
invariance, i.e., Ward identities, dictate the graviton-fermion
interactions. The equivalence principle is a consequence of gauge
invariance and remains as true as ever. Because of gauge invariance, the
definition of length and time measurement must involve the gravitational
field, ultimately to the same effect as if operating in curved
space-time$^{2,3}$.

\section{Non-Perturbative Solutions.}
The question is now to what extent non-perturbative results remain true
in the dimensional approach. Let us be clear about the extent to which
non-perturbative results in Hamiltonian or path integral formalism
have been derived and used. By necessity, since radiative corrections
need regularization, most (but not all) applications are based on the
tree level approximation. That produces classical electromagnetism
and the general theory of relativity. There may or may not be differences
with the dimensional formulation, depending on whether some extension of
perturbation theory may be made plausible. Furthermore, path integrals
have been used as tools for numerical calculations, but also for
certain theoretical considerations. It is hard to say whether this
proves anything one way or the other. It would be quite interesting
if the gap between perturbation theory and the strong coupling limit
within the path integral approach could be bridged. In short, while
many blindly accept the validity of the Hamiltonian or path integral
formalism, there is really no objective basis on which to accept the
validity in the non-perturbative region. This in addition to all kinds
of problems in the case of path integrals with respect to chiral fermions.

Various different situations must be envisaged. We first turn to ordinary
bound states such as the atom. Do such solutions survive in a perturbative
approach?

As is well known the Schr\"odinger, or rather the Lippmann-Schwinger
equation, can be derived from diagram theory by means of a partial sum
of diagrams. In this case the diagrams to be summed are the ladder
diagrams. In the approximation of low momentum transfer the
Lippmann-Schwinger equation for situations such as electron proton
scattering can be derived. Bound states can be understood from the
analytic continuation of the amplitude to negative energies, and these
states manifest themselves as poles in the complex energy plane along
the negative real axis. This type of analytic continuation is analogous
to that encountered in connection with unstable particles. Also there
one has a perturbation series that diverges in a particular momentum
region; there the Dyson summation provides us with a solution whose
perturbation expansion coincides with perturbation theory wherever that
expansion converges, and which is otherwise an analytic continuation to
the region where perturbation theory fails. We will accept such
solutions as quasi-perturbative solutions. In that sense then the more
obvious non-perturbative situations such as atoms and planetary systems
can still be understood from the perturbative point of view. And nuclei
would also fall in this class.

The situation becomes more difficult with respect to other solutions of
the field theoretical equations of motion. Usually one writes formally
equations of motion in coordinate space, and in the dimensional approach
we cannot accept those solutions unless they somehow can be understood
from the point of view of momentum space. The above discussion
examplifies that. There is now one type of argument that is completely
beyond any momentum space formulation, and that is the treatment of a
$\theta$ term as commonly introduced to derive $CP$ violating effects in
quantum chromodynamics$^4$. Such a term is a total derivative and as
such has no counterpart in momentum space. The instanton vacuum with its
winding number requires the concept of absolute space and boundary
conditions to that space. From a perturbative point of view, with space
defined mathematically through Fourier transformation, such solutions
are incomprehensible. In general, assigning absolute properties to the
vacuum in coordinate space is meaningless in momentum space. In the
perturbative approach we must reject such constructs as based
essentially on the prejudice of an absolute space. Thus in the
dimensional formulation there is no strong $CP$ problem as there is no
such thing as the instanton vacuum.

\section{Dimensional Regularization.}
At this time we do not want to go into technical details concerning
dimensional regularization, but feel nonetheless compelled to state
that there are some non-trivial problems there that have not yet been
settled satisfactorily, at least in our opinion. There is of course
the usual problem of treating $\gamma^5$; it is our understanding that
this problem is essentially understood and that there is no difficulty
there$^5$. Here we will not enter into any discussion on that, and neither
will we consider related issues such as chiral invariance. There are
however further problems in case of fermion lines that end in external
lines, and that may also be part of closed loops. Then it is not clear
how $\gamma$'s are to be treated, since one is dealing with a string
terminated by spinors that have no trivial generalization to
continuous $n$. We refer to the literature for a more extensive
discussion of this problem$^6$.

Another problem that arises concerns the $CPT$ theorem. In spinor space
the transformation matrix corresponding to the $CPT$ transformation is
$\gamma^5$, and we may have a problem. Stated otherwise, in $n$
dimensions the $PT$ transformation involves only the first four
components of vector quantities and not those beyond the fourth
dimension. Thus components of loop momenta beyond the fourth behave
anomalously under $CPT$. However, the deviations arising from that will
go to zero in the limit $n=4$, and appear not to result in observable
consequences if all singularities at $n=4$ have been subtracted
properly\footnote{$^*$}{\ninepoint$\!\!\!\!\!$ In case of anomalies there is
seepage out of four dimensional space due to an unsubtracted pole.}.
In a non-renormalizable environment such as gravitation
there would result finite $CPT$ violating effects, but we can hardly
take that serious at this point.

Let us close this section with a remark. One might think that $CPT$ in
$n$ dimensions can be defined as reversal of all coordinates. But that
poses a problem: going from odd to even dimension the determinant of the
$PT$ transformation changes sign, and in odd dimensions the $PT$
transformation is not continuously related to the identity. In any case,
in the usual formulation, in four dimensions, the $CPT$ transformation
is represented in spinor space by $\gamma^5$, and the problem of
generalizing $\gamma^5$ is known to all. Now perhaps this can be
remedied by limiting oneself to transformations that involve only one of
the spatial-like coordinates and in addition the time-like coordinate,
but we will not discuss that here.

\section{Naturalness, Supersymmetry and the Mass Equation.}
In a previous publication$^7$ we have argued for a mass equation
resulting from the requirement that there be no quadratic divergences.
The meaning of a quadratic divergence is completely unclear within the
method of dimensional regularization, and the treatment offered was
based on a vague analogy between the dimensional method and a momentum
cut-off scheme.

At this point we would like to distance ourselves from such an
approach. Quadratic divergencies do not exist within the dimensional
formulation. The concept of naturalness with respect to scalar particle
masses needs revision. There are no large corrections related to
quadratic divergencies as these divergencies do not exist in the
dimensional method. Of course, corrections to scalar particle masses
involving masses of heavier particles could still occur, but that is a
quite different subject. Only within a well defined model can
conclusions be drawn.

Supersymmetry has evolved on the premise that this solves the
naturalness problem with respect to the Higgs mass. That is really not a
very strong argument if we realize that this requires the idea that
somehow quadratic divergencies become finite through some physical
cut-off mechanism, and that there is a scale associated with that. The
singular non-success of supersymmetry so far supports, and in fact to
some extent produced our negative view. Within the dimensional approach
it is simply not clear what purpose would be served by a supersymmetric
theory. That is no proof against a possible existence, but it certainly
weakens the case.

The question is if there is any equation left relating the top and Higgs
mass. Very speculatively we would like to argue that tadpole type
diagrams should add up to zero in view of difficulties with the
cosmological constant. In lowest order that results in the same equation
as before. If both top mass $m_t$ and Higgs mass $m_H$ are large with
respect to all other masses the relation is roughly $m_H = 2 m_t$. We
emphasize the highly speculative nature of this relation.

\section{Gravitation.}
The theory of gravitation can be formulated as a gauge theory, and that
is of course precisely as it is always done in quantum field theory. The
trouble is that the theory is non-renormalizable, and the dimensional
scheme offers no new insight here. Thus we are facing an unsatisfactory
theory from the start, no matter what starting point. Nonetheless we may
perhaps spend some words on the question of black holes and the problem
of the cosmological constant. The arguments may well be naive and
incomplete.

In the traditional formulation of gravitation, black holes are
an unavoidable consequence of Einstein's equations. In the dimensional
formulation, lacking coordinate space, we may ask to what extent such
solutions remain valid. Can they somehow be seen as analytical
continuations of solutions valid in a perturbative domain, like the
bound states discussed above?

First, the approximation in terms of ladder diagrams, so succesful in
understanding the hydrogen atom. Here there is a peculiar difficulty
with gauge invariance: ladder diagrams do not form a gauge invariant
set. In the case of electromagnetism one can include crossed ladders,
and then the result is gauge invariant. However, in the case of
gravitation the approximation is never gauge invariant because of
gravitational self-coupling. So, even to explain planetary systems, one
must somehow approximate further, and we leave it to the reader to
realize precisely the approximations involved in the standard classical
approach.

However, as a consequence we have no description of the bound state
problem that can be extended or analytically continued to the case that
the gravitational self-coupling becomes important.

In terms of the gravitational field itself we may consider the solutions
of the classical equations of general relativity. For a Schwarzschild
black hole with radius $R$ the spatial components of the gravitational
tensor in cartesian coordinates are $h_{jk}={x_jx_k \over r^2} {R \over
r-R}$. The Fourier transform of this is non-existent, and also cannot be
defined as a function of $R$ in some region and then continued to the
region of positive real $R$.

The arguments presented here are certainly not complete. The fields
$h_{\mu\nu}$ are not gauge invariant, and perhaps there is a choice of
gauge (choice of coordinates) in which the Fourier transform exists or
can be defined in some way. Nonetheless it is tempting to deny the
existence of black holes, and in any case, it must be realized what the
underlying assumptions are in the traditional approach. It might be
added that the remarkable absence of black holes outside the domain of
astrophysical speculation tends to support the idea of relative space
and the perturbative approach.

Concerning the cosmological constant problem, at first sight one might
think that the problem is non-existent. Given that there is no such
thing as the coordinate space vacuum there is no way of assigning
properties such as curvature to that. Unfortunately the problem surfaces
in a different form.

The gravitational Lagrangian is of the form:
$${\cal L}_{grav} = -\sqrt{g}\, \left(R + \lambda\right)\, ,$$ where $g$
is the determinant of the metric tensor $g_{\mu\nu}$ and $R$ is the
Riemann scalar. The quantity $\lambda$ is the cosmological constant, and
radiative corrections affect it. When expanding $g_{\mu\nu} =
\delta_{\mu\nu} + h_{\mu\nu}$, where $h_{\mu\nu}$ is the gravitational
field, a tadpole type term $\lambda h_\mu^\mu$ arises. The conventional
field theoretical treatment of terms linear in the field is to perform a
shift, $h_{\mu\nu} \to h_{\mu\nu} + a_{\mu\nu}$, choosing $a_{\mu\nu}$
such that the linear term disappears. Conventionally, as for example in
the $\sigma$-model, the quantity $a$ is a constant, but in the case of
gravity the solution is space-time dependent, that is $a_{\mu\nu} =
a_{\mu\nu}(x)$. The equation arising from the condition that no linear
term arises is precisely the Einstein equation for the case of no
matter; in more conventional terms of space-time it is the solution
corresponding to a curved universe. One might say that there is a very
strong background gravitational field (the field $a_{\mu\nu}(x)$)
present in the vacuum, and physical processes evolve against this
background.

In the perturbative perspective this is all but transparant. Our main
problem is with unitarity. The $S$-matrix arising when considering the
theory with a momentum dependent shift of the gravitational field is not
evidently unitary. There are then vertices coupling particles to a
classical source (the $a_{\mu\nu}$ field). It would mean spontaneous
creation or absorbtion of particles. We are at a loss to make any sense
out of this. In other words, a non-zero cosmological constant appears
inconsistent within the perturbative approach, but there is no way to
guarantee its vanishing.

Assigning absolute meaning to time, we could discuss questions
concerning the beginning of the universe. At this point, given the
unsatisfactory situation concerning gravitation we believe such a
discussion to be premature.

\section{Conclusions}The basic assumptions that may be used as a
starting point for the description of physical reality are not
equivalent. Space and time are not part of the perturbative dimensional
formulation and are thus defined only through Fourier transformation.
Many non-perturbative results of contemporary field theory may be
questioned, and we have attempted to differentiate between solutions
that can be obtained by some analytic continuation and those that have
no connection whatsoever to perturbation theory.

The question may be raised if the issues discussed here can ever be
resolved. The non-success of non-perturbative solutions may be
considered circumstantial evidence against coordinate space
formulations, but is no proof. If we do find however any phenomenon that
is particular to the dimensional formulation then that could decide the
issue. But then it may well be that that is not the ultimate formulation
either. There certainly is need for improvement.

\section{Acknowledgements}The author would like to thank Professors R.
Akhoury, W. Ford and V. Zakharov and Dr H. Veltman for critical remarks
and constructive comments.

\vfil\eject

\references

\item{1.} A. Einstein, The Meaning of Relativity, edition 1953,
 especially \S 5, page 161.
\item{2.} R.P. Feynman, Lectures on Gravitation,
 California Institute of Technology lecture notes 1962-63.
\item{3.} M. Veltman, Quantum theory of gravitation. Les Houches
 lectures 1975.
\item{4.} V. Baluni, Phys. Rev. {\bf D19} (1979) 2227.\hfil\break
 See also for further references R.J. Crewther, P. Di Vecchia,
 G. Veneziano and E. Witten, Phys. Let. {\bf 88B} (1979) 123.
\item{5.} P. Breitenlohner and D. Maison,
 Comm. Math. Phys. {\bf 52} (1977) 11.\hfil\break
 C. Becchi, A. Rouet and R. Stora,
  Annals of Physics {\bf 98} (1976) 287.\hfil\break
 B. Bandelloni, A. Blasi, C. Becchi and R. Collina,
 Ann. Inst. Henri Poincar\'e, {\bf 28} (1978) 225.
\item{6.} K. Adel, McGill preprint March 1994.
\item{7.} M. Veltman, Acta Physica Polonica {\bf B12} (1981) 437.

\end